\documentstyle[12pt,epsf,rotate,a4]{article}

\def\del{{\partial}}
\def\al{{\alpha}}
\def\be{{\beta}}
\renewcommand{\theequation}{\arabic{section}.\arabic{equation}}

\begin{document}

\baselineskip=24pt

\title{
\vspace{-3.0cm}
\begin{flushright}  
{\normalsize September, 1996}\\
\vspace{-0.3cm}
{\normalsize UTHEP-345}\\
\vspace{-0.4cm}
{\normalsize hep-th/9609163}\\
\end{flushright}
\vspace*{2.0cm}
{\center{\Large {\bf Mirror Symmetry of Calabi-Yau Manifolds \\
and \\
Flat Coordinates}}}
\bigskip
\bigskip
\bigskip
\bigskip
       }

\author{ Masayuki Noguchi\thanks{e-mail: noguchi@het.ph.tsukuba.ac.jp}
 \\ \bigskip \\
        \it{Institute of Physics,
            University of Tsukuba,} \\ \it{Ibaraki 305, Japan}}
\date{}

\maketitle

\vspace*{0.5cm}

\begin{abstract}

We study mirror symmetry of Calabi-Yau manifolds
within the framework of the Gauss-Manin system.
Applying the flat coordinates to the 
Gauss-Manin system for the periods,
we derive differential equations for the 
mirror map in addition to the ordinary Picard-Fuchs
equations for the periods.
These equations are obtained for a class of one-parameter models
and a two-parameter model of Fermat type 
Calabi-Yau manifolds.

\end{abstract}

\newpage
\setcounter{equation}{0}

\section{Introduction}

In string theory, one of the most important problems is to derive
a realistic low-energy theory. To discuss low-energy theories,
one has to determine masses, gauge coupling constants, etc.
These various couplings depend on
the moduli of the string model. In the context of conformal
field theory, the moduli are described by the marginal operators.
When we consider the Calabi-Yau compactifications
down to four dimensions, these are identified as
the harmonic $(1,1)$ and $(2,1)$ forms which describe the K\"{a}hler and
complex structure deformations, respectively. The number of the
deformation parameters is equal to the Hodge numbers $h_{1,1}$ and $h_{2,1}$.
Since there are no couplings between the two different types of 
deformations, 
the Yukawa couplings are split into two sectors. 
It is well-known that the Yukawa 
couplings corresponding
to complex structure deformations receive no loop collections. 
On the other hand, the Yukawa couplings
corresponding to K\"{a}hler deformations 
are renormalized by instantons
\cite{W} \cite{Ma} \cite{DS}.
Therefore at first glance 
one can analyze half of the moduli dependence in the 
Calabi-Yau compactifications exactly.

It was, however, discovered by Candelas et al.$\ $\cite{C} that 
a mirror symmetry of the Calabi-Yau manifolds enables us to determine
the full dependence of Yukawa couplings on the moduli. A 
mirror symmetry relates two topologically distinct 
Calabi-Yau manifolds $M$ and $\bar M$ which have the relations
$h_{1,1}=\bar{h}_{2,1}$ and $h_{2,1} = \bar{h}_{1,1}$
with $h$ and $\bar h$ being the Hodge numbers of $M$ and $\bar M$,
respectively.
A more important property of this symmetry is to give 
isomorphism between the quantum cohomology ring of $M$ and $\bar M$
\cite{GP}. So,
by virtue of this symmetry, one can derive the exact 
dependence of Yukawa couplings on 
K\"{a}hler deformations from tree level informations of 
mirror manifold \cite{C} \cite{M} \cite{F} \cite{KT} \cite{HKTY}
\cite{COFKM} \cite{CFKM}. 
In these calculations, the crucial point is to  construct
the mirror map explicitly
which relates a mirror pair of Calabi-Yau manifolds.
Usually this is done by making a monodromy analysis 
of periods in the large radius limit.

Recently, Hosono and Lian have clarified
that the mirror map is intimately related to the 
notion of flat coordinates of 
the period integrals \cite{HL}. The idea of the flat coordinates \cite{S}
has found extensive applications, 
for example, off-critical deformations of $N=2$
superconformal theories in $2D$
\cite{CV2}, $2D$ quantum gravity and topological field theories
\cite{Wqg} \cite{DVV} \cite{BV} \cite{VW}
\cite{LSW} \cite{KTS} \cite{EYY}. 

In the context of mirror symmetry of Calabi-Yau manifolds, however, it was
tantalizing to see how the flat coordinates play a role 
when the mirror map is considered. In \cite{HL},
taking the quintic Calabi-Yau manifold as an example,
the flat coordinates are introduced to define a natural
basis of the quantum cohomology ring. It is then 
observed that the flat coordinates are related to 
the solution of the Picard-Fuchs equations at the point
of maximally unipotent monodromy. All this is done in the 
framework of the Gauss-Manin system 
which is a set of the first order differential equations
for the period integrals.

 In this article, we extend the 
analysis of \cite{HL} by taking more examples of 
Calabi-Yau manifolds with one K\"{a}hler modulus
and two K\"{a}hler moduli. It will be shown
that differential equations which govern the 
periods and the mirror map are naturally obtained by applying 
the flat coordinates to the Gauss-Manin system.

In section 2, employing the Gauss-Manin differential equations in the 
flat coordinate system, we analyze a class of 
one-parameter models which have been studied in \cite{KT}.
In section 3, the flat coordinate method is applied to a two-parameter 
model. We study the model defined on the 
weighted projective space ${\bf P}(1,1,2,2,2)$ whose mirror map has 
been investigated in \cite{HKTY} \cite{COFKM}. Finally,
section 4 is devoted to our conclusions.

\setcounter{equation}{0}
\section{Flat coordinates for one-parameter models}

We first consider the simplest Calabi-Yau manifolds
which have the Fermat type defining relation such that
$W = \sum_{i=0}^4 x_i^{n_i}$
with a one-parameter deformation of 
K\"ahler class, i.e. $h_{1,1}=1$.
As is listed in \cite{KT}, these models are characterized by
the integer $k$ with $k=5,6,8$ and $10$. 
The defining relations are given by
\begin{equation}
\begin{array}{l}
k=5 :\cr 
M = \{ x_i \in {\bf P}(1,1,1,1,1)|
W_5 = {1\over 5}(x_0^5 + x_1^5 + x_2^5 + x_3^5 + x_4^5) = 0 \},\cr
k=6 :\cr
M = \{ x_i \in {\bf P}(2,1,1,1,1)|
W_6 = {1\over 3} x_0^3 + {1\over 6} (x_1^6 + x_2^6 + x_3^6 + x_4^6) = 0 \},\cr
k=8 : \cr
M = \{ x_i \in {\bf P}(4,1,1,1,1)|
W_8 = {1\over 2} x_0^2 + {1\over 8} (x_1^8 + x_2^8 + x_3^8 + x_4^8) = 0 \},\cr
k=10 :\cr
M = \{ x_i \in {\bf P}(5,2,1,1,1)|
W_{10} = {1\over 2} x_0^2 + {1\over 5} x_1^5 + {1\over 10} 
(x_2^{10} + x_3^{10} + x_4^{10}) = 0 \}.
\label{cy1}
\end{array}
\end{equation}
These models have first appeared in \cite{SW} and 
possess the Hodge numbers $h_{1,1}=1$ and 
$h_{2,1}=101,103,149$ and $145$ for $k=5,6,8$ and $10$, respectively.

To get the mirror manifolds, one 
must consider the orbifoldization of these models
by dividing out a certain discrete
symmetry $G$ of the defining relation.
There is a deformation of the
defining relations (\ref{cy1}) that preserves invariance under this
discrete symmetry. We have the deformed polynomials
\begin{equation}
W_k(\al) = W_k - \al x_0 x_1 x_2 x_3 x_4,
\end{equation}
where $\al$ is a deformation parameter.
Let $M_{\al}$ be the manifold defined by $W_{k}(\al)=0$,
then its mirror manifold $\bar{M}_{\al}$ is obtained as
\begin{equation}
\bar{M}_\al = M_\al/G,
\end{equation} 
where 
$G$ is ${\bf Z}_5^3, \ {\bf Z}_3 \times {\bf Z}_6^2,  \ {\bf Z}_8^2 
\times {\bf Z}_2$ and ${\bf Z}_{10}^2$
for $k = 5,6,8$ and $10$, respectively. 
The deformation by $\al$ 
can now be regarded as the 
complex structure deformation of the orbifold (mirror manifold).
Actually, by direct calculations
on the manifolds \cite{KT}, 
one can see that these
manifolds $\bar{M}_\al$ have the Hodge numbers 
$h_{1,1} = 101,103,149,145$ and $h_{2,1} = 1$,
and hence these are mirrors of $M$'s.

In order to discuss the quantum cohomology ring of ${\bar M}_{\al}$ using
the Gauss-Manin system,
we introduce the deformed Jacobian ring $J_{\al}$ by
\begin{equation}
J_\al \equiv {\bf{C}}[x_0,\dots,x_4]^G/(\del W_k(\al)/\del x_i)\cap 
{\bf C}[x_0,\dots,x_4]^G.
\label{djr}
\end{equation}
We fix a basis of $J_\al$ as $\{1,\phi,\phi^2,\phi^3\}$
where $\phi \equiv x_0 x_1 x_2 x_3 x_4$.
Let us define the period integrals $w^{(i)}_{\gamma_j}$ as
\begin{equation}
w_{\gamma_j}^{(i)} = i!
\int_{\gamma_j} {\phi^i \over W_k^{i +1}(\alpha)}
d\mu ,
\ \ \
i,j=0,1,2,3,
\label{pi}
\end{equation}
where 
$d\mu = \sum_{l=0}^4 (-1)^{l+1} x_l dx_0 \wedge \cdots \wedge \hat{dx_l}
\wedge \cdots \wedge dx_4$ and $\gamma_j$'s are the homology cycles in 
$H_3(\bar{M}_\al,{\bf Z})$.
Using the relations obtained from $\del W_\al/ \del x_i = \dots$ and 
integrating by parts,
we derive a set of first order 
differential equations which are satisfied by the periods (\ref{pi}) 
\begin{equation}
{\del \over \del \alpha}w = G_k(\alpha)w,
\label{gm}
\end{equation}
where $w =\  ^t (w^{(0)}_{\gamma_j},w^{(1)}_{\gamma_j}
                ,w^{(2)}_{\gamma_j},w^{(3)}_{\gamma_j})$ and 
\begin{eqnarray}
G_5(\alpha)&=& \left(
\begin{array}{cccc}
0&1&0&0 \\
0&0&1&0 \\
0&0&0&1 \\
{\alpha \over 1-\alpha^5}&{15\alpha^2 \over 1-\alpha^5}&
{25 \alpha^3 \over 1-\alpha^5}&{10 \alpha^4 \over 1-\alpha^5}
\end{array}
\right),\cr
G_6(\alpha) &=& \left(
\begin{array}{cccc}
0&1&0&0 \\
0&0&1&0 \\
0&0&0&1 \\
{\alpha^2 \over 1-\alpha^6}&{15\alpha^3 \over 1-\alpha^6}&
{25 \alpha^6 - 2 \over \alpha^2 (1-\alpha^6)}&{10 \alpha^6 + 2
 \over \alpha (1-\alpha^6)}
\end{array}
\right), \cr
G_8(\alpha) &=& \left(
\begin{array}{cccc}
0&1&0&0 \\
0&0&1&0 \\
0&0&0&1 \\
{\alpha^4 \over 1-\alpha^8}&{15(\alpha^8 + 1) \over \al^3(1-\alpha^8)}&
{5 (5 \alpha^8 - 3)\over \al^2 (1-\alpha^8)}&{2 (5\al^8 + 3)
 \over \al (1-\alpha^8)}
\end{array}
\right),\cr
G_{10}(\alpha) &=& \left(
\begin{array}{cccc}
0&1&0&0 \\
0&0&1&0 \\
0&0&0&1 \\
{\alpha^6 \over 1-\alpha^{10}}&{5(3\alpha^{10} + 7) 
\over \al^3(1-\alpha^{10})}&
{5 (5 \alpha^{10} - 7)\over \al^2 (1-\alpha^{10})}&{10 (\al^{10} + 1)
 \over \al (1-\alpha^{10})}
\end{array}
\right).
\label{gk}
\end{eqnarray}
These are the Gauss-Manin system for the periods.
These first order systems reduce to the ordinary
differential equations for $w^{(0)}_{\gamma_j}$
which take the forms
\begin{eqnarray}
k = 5&:& (1 - \al^5) {w^{(0)}}'''' - 10 \al^4 {w^{(0)}}''' 
- 25 \al^3 {w^{(0)}}'' 
- 15 \al^2 {w^{(0)}}' - \al w^{(0)} = 0,\cr
k = 6&:&
\al^2 (1 - \al^6) {w^{(0)}}'''' - 2 \al (5 \al^6 + 1){w^{(0)}}'''  \cr
 & &  
 \ \ \ - (25 \al^6 - 2) {w^{(0)}}'' - 15 \al^5 {w^{(0)}}' - \al^4 {w^{(0)}}
 = 0, \cr
k = 8&:&
\al^3 (1 - \al^8) {w^{(0)}}'''' - 2 \al^2 (5 \al^8 + 3) {w^{(0)}}'''  \cr
 & & 
\ \ \ - 5 \al (5 \al^8 - 3) {w^{(0)}}'' - 15 (\al^8 + 1) {w^{(0)}}' 
- \al^7 {w^{(0)}}
=0, \cr
k = 10&:&
\al^3 (1 - \al^{10}){w^{(0)}}'''' - 10 \al^2 (\al^{10} + 1) {w^{(0)}}'''  \cr
 & & 
\ \ \ - 5 \al (5 \al^{10} - 7){w^{(0)}}'' - 5 (3 \al^{10} + 7){w^{(0)}}'
 - \al^9 w^{(0)} = 0
\label{pf}
\end{eqnarray}
where $' = \del / \del \al$. 
These are nothing but the Picard-Fuchs equations \cite{KT}.

Let us now introduce the flat coordinates and rewrite the 
Gauss-Manin system.
For this we  first make
a degree-preserving change of basis of the Jacobian ring 
\begin{equation}
\{1,\phi,\phi^2,\phi^3\} \rightarrow 
\{{\cal O}^{(0)},{\cal O}_t^{(1)},{\cal O}^{t(2)},{\cal O}^{(3)}\}
\end{equation}
with ${\cal O}^{(0)} = 1$. Here we 
regard $\al$ as a function of the flat coordinate $t$ ($\al = \al(t)$).
For this basis, the period integrals are defined by
\begin{equation}
v^{(i)}_{\gamma_j} = (-1)^{i}i! \int_{\gamma_j} 
{{\cal O}^{(i)} \over (r_{11} W_k(\al))^{i+1}} d\mu,
\label{fcpi}
\end{equation}
where we have changed the normalization of the defining
relation by a certain function $r_{11}(\al)$. 
The period integrals (\ref{fcpi}) are related to
(\ref{pi}) through
\begin{equation}
w(\al) = M_k(\al)v(t),
\label{210}
\end{equation}
where $w(\al) = {}^t(w^{(0)},w^{(1)},w^{(2)},w^{(3)})$,
$v(t) = {}^t(v^{(0)},v^{(1)},v^{(2)},v^{(3)})$ and 
$M_k (\al)$ is a lower-triangular matrix 
\begin{equation}
M_k(\al) = \left(
\begin{array}{cccc}
r_{11}&0&0&0 \\
r_{21}&r_{22}&0&0 \\
r_{31}&r_{32}&r_{33}&0 \\
r_{41}&r_{42}&r_{43}&r_{44}
\end{array}
\right).
\end{equation}
Substituting (\ref{210}) into (\ref{gm})
and $\al \rightarrow t=t(\al)$, we see that 
the Gauss-Manin system for $v$ becomes
\begin{equation}
{\del \over \del t} v = \left( {\del \al \over \del t}\right)
\left(M_k^{-1} G_k M_k - 
M_k^{-1} {\del M_k \over \del \al } \right) v.
\label{star}
\end{equation}

The basis 
$\{{\cal O}^{(0)},{\cal O}_t^{(1)},{\cal O}^{t(2)},{\cal O}^{(3)}\}$
is constrained to satisfy 
the flat coordinate conditions
\begin{eqnarray}
{\rm i)}& &{\cal O}_t^{(1)} = {\del \over \del t}(r_{11} W), \cr
{\rm ii)}& &{\cal O}_t^{(1)} {\cal O}_t^{(1)} = K_{ttt}(t) {\cal O}^{t(2)}, \cr
{\rm iii)}& &{\cal O}_t^{(1)} {\cal O}^{t(2)} = {\cal O}^{(3)}, \cr
{\rm iv)}& &{\cal O} {\cal O}^{(3)} = 0,\ \ \ 
({\cal O} = {\cal O}_t^{(1)},{\cal O}^{t(2)}).
\end{eqnarray}
Then we have
\begin{equation}
{\del \over \del t} v
=
\left(
{\begin{array}{cccc}
0&1&0&0 \\
0&0&K_{ttt}(t)&0 \\
0&0&0&1 \\
0&0&0&0
\end{array}}
\right) v.
\label{fcgm}
\end{equation}
Comparing (\ref{star}) with (\ref{fcgm})
we determine the transformation matrix $M_k$.
After some algebra we obtain
\begin{equation}
\begin{array}{l}
M_5 = \left(
\begin{array}{cccc}
r_{11}&0&0&0 \\
r_{11}'&r_{11} t'&0&0 \\
r_{11}''&r_{11} t''+2 r_{11}' t'&{1 \over r_{11} t'}
{C \over 1 - \al^5}&0 \\
r_{11}'''&r_{11}t''' + 3 r_{11}' t'' + 3 r_{11}'' t'&
{C\over r_{11} t'}\{({1\over 1-\al^5})'+
{r_{11}'\over r_{11}}{1\over 1-\al^5}\} & 
{1\over r_{11}}{C\over 1-\al^5}
\end{array}
\right), 
\cr
M_6 = \left(
\begin{array}{cccc}
r_{11}&0&0&0 \\
r_{11}'&r_{11} t'&0&0 \\
r_{11}''&r_{11} t''+ 2 r_{11}' t'&{1 \over r_{11} t'}
{C\al \over 1 - \al^6}&0 \\
r_{11}'''&r_{11} t''' + 3 t_{11}' t'' + 3 r_{11}'' t'&
{C\over r_{11} t'}\{({\al\over 1-\al^6})'+
{r_{11}'\over r_{11}}{\al\over 1-\al^6}\} &
{1\over r_{11}}{C\al\over 1-\al^6}
\end{array}
\right), 
\cr
M_8 = \left(
\begin{array}{cccc}
r_{11}&0&0&0 \\
r_{11}'&r_{11} t'&0&0 \\
r_{11}''&r_{11} t' + 2 r_{11}' t'&{1 \over r_{11} t'}
{C\al^3 \over 1 - \al^8}&0 \\
r_{11}'''&r_{11} t''' + 3 r_{11}' t'' + 3 r_{11}'' t'&
{C\over r_{11} t'}\{({\al^3\over 1-\al^8})'+
{r_{11}'\over r_{11}}{\al^3\over 1-\al^8}\} &
{1\over r_{11}}{C\al^3\over 1-\al^8}
\end{array}
\right), 
\cr
M_{10} = \left(
\begin{array}{cccc}
r_{11}&0&0&0 \\
r_{11}'&r_{11} t'&0&0 \\
r_{11}''&r_{11} t'' + 2 r_{11}' t'&{1 \over r_{11} t'}
{C\al^5 \over 1 - \al^{10}}&0 \\
r_{11}'''&r_{11} t''' + 3 r_{11}' t'' + 3 r_{11}'' t'&
{C\over r_{11} t'}\{({\al^5\over 1-\al^{10}})'+
{r_{11}'\over r_{11}}{\al^5\over 1-\al^{10}}\} &
{1\over r_{11}}{C\al^5\over 1-\al^{10}}
\end{array}
\right),
\label{tmat}
\end{array}
\end{equation}
where $r_{11}(\al)$ and $t(\al)$ satisfy the following 
differential equations 
\begin{eqnarray}
k=5:&
r_{11}'''' - {10\al^4\over 1-\al^5}r_{11}''' - {25\al^3\over 1-\al^5}r_{11}''
- {15\al^2\over 1-\al^5}r_{11}' - {\al \over 1-\al^5}r_{11} = 0, \cr
& t''' - \left({5\al^4 \over 1 - \al^5} - {2 r_{11}'\over r_{11}} \right) t''
- \left({5\al^3 \over 1 - \al^5} + 
{10 \al^4 \over 1 - \al^5}{r_{11}'\over r_{11}} +
{2 r_{11}'^2 \over r_{11}^2} - {4 r_{11}'' \over r_{11}} \right) t' = 0,\cr
k=6:&
r_{11}'''' - {2(5\al^6+1)\over \al(1-\al^6)}r_{11}''' -
{(25\al^6 -2)\over \al^2(1-\al^6)}r_{11}'' - {15\al^3\over 1-\al^6}r_{11}'
-{\al^2\over 1-\al^6}r_{11} =0, \cr
& t''' - \left({5\al^6 + 1 \over \al (1 - \al^6)} - 
{2 r_{11}'\over r_{11}} \right) t'' -
\left({5 \al^4 \over 1 - \al^6} + 
{10 \al^6 + 2 \over \al(1 - \al^6)}{r_{11}'\over r_{11}} 
+ {2 r_{11}'^2 \over r_{11}^2} - {4 r_{11}''\over r_{11}} \right) t' = 0,\cr
k=8:&
r_{11}'''' - {2(5\al^8+3)\over \al(1-\al^8)}r_{11}''' -
{5(5\al^8-3)\over \al^2(1-\al^8)}r_{11}'' - 
{15(\al^8+1)\over \al^3(1-\al^8)}r_{11}' - {\al^4\over 1-\al^8}=0,\cr
&t''' - \left({5\al^8 + 3 \over \al(1 - \al^8)} - 
{2 r_{11}'\over r_{11}}\right)t'' -
\left( {5\al^8 - 3\over \al^2(1 - \al^8)} + 
{10 \al^8 + 6 \over \al(1 - \al^8)}{r_{11}'\over r_{11}} +
{2 r_{11}'^2 \over r_{11}^2} - {4 r_{11}'' \over r_{11}}\right) t'=0, \cr
k=10:&
r_{11}'''' - {10(\al^{10}+1)\over \al(1-\al^{10})}r_{11}'''-
{5(5\al^{10}-7)\over \al^2(1-\al^{10})}r_{11}''-
{5(3\al^{10}+7)\over \al^3 (1-\al^{10})}r_{11}'-
{\al^6\over 1-\al^{10}}r_{11}=0, \cr
&t''' - \left( {5\al^{10} + 5 \over \al(1 - \al^{10})} - 
{2 r_{11}'\over r_{11}} \right)t'' - \left(
 - {5 \over \al^2} + 
{10\al^{10} + 10 \over \al(1 - \al^{10})}{r_{11}'\over r_{11}} +
{2 r_{11}'^2 \over r_{11}^2} - {4 r_{11}''\over r_{11}} \right)t' = 0.
\label{def}
\end{eqnarray}
The first equation for each $k$ coincides with the Picard-Fuchs equation
(\ref{pf}),
and the second equation determines the flat coordinate $t$.
Moreover, from eqs.(\ref{star})$\sim$(\ref{tmat}),
the Yukawa couplings are obtained as
\begin{equation}
K_{ttt} = {1\over {r_{11}}^2} {C\al^{k-5} \over 1 - \al^k }
\left( {\del \al \over \del t}\right)^3, \ \ \ k = 5,6,8,10,
\label{yc}
\end{equation}
where $C$ is a constant. This is a desired form in agreement with
\cite{C} \cite{KT}.

The power series solution of the Picard-Fuchs equations (\ref{pf})
around $\al = \infty$ can be found easily.
The solution is given by
\begin{equation}
\bar{w}_0^{(0)}(\al) = {1\over \al} \sum^\infty_{m=0}
{\Gamma(km+1)\over \prod^4_{i=0} \Gamma(\nu_i m+1)}
(\gamma \al)^{-k m}, \ \ |\al|>1,
\label{sol}
\end{equation}
where 
\begin{equation}
{\bf \nu} = \left\{
\begin{array}{lcl}
(1,1,1,1,1)&:&k=5 \cr
(2,1,1,1,1)&:&k=6 \cr
(4,1,1,1,1)&:&k=8 \cr
(5,2,1,1,1)&:&k=10
\end{array}
\right.
\end{equation}
and $\gamma = k \prod^4_{i=0} (\nu_i)^{-\nu_i/k}$.
Other solutions ${\bar w}_j^{(0)}(\al)$
are generated from (\ref{sol}) by
\begin{equation}
\bar{w}_j^{(0)}(\al) = \bar{w}_0^{(0)}(\delta^j \al), 
\ \ \ \delta = e^{2\pi i/k}.
\end{equation}
As a set of fundamental solutions Klemm and Theisen 
have chosen ${\bar w}_j^{(0)}$ with
$j=0,1,2$ and $k-1$ \cite{KT}.

On the other hand, it is difficult to deal with
the second equations obeyed by $t$. 
Let us first examine the asymptotic behavior
of $t$. In the limit $\al \rightarrow \infty$, 
eqs.(\ref{def}) turn out to be
\[
r_{11}'''' + {10 \over \al} r_{11}''' + {25 \over {\al^2}} r_{11}''
+ {15 \over {\al^3}} r_{11}' + {1 \over {\al^4}} r_{11} = 0,
\]
\begin{equation}
t''' + 
\left( {5 \over \al} + {2 r_{11}' \over r_{11}}
\right) t'' +
\left( {5 \over \al^2} + {10 \over \al} {r_{11}' \over r_{11}} 
      -{2 r_{11}'^2 \over r_{11}^2} + {4 r_{11}'' \over r_{11}}
\right) t' = 0
\end{equation}
for $k=5,6,8$ and $10$. The solutions of the first equation reads
\begin{equation}
r_{11} \sim {1\over \al},\ \ {{\rm log}\al \over \al}, \ \
         {({\rm log}\al)^2 \over \al},\ \ {({\rm log}\al)^3 \over \al}.
\end{equation}
If we choose the solution $r_{11} \sim 1/\al$,
then the asymptotic solutions of $t$ are
\begin{equation}
t \sim {{\rm log}\al},\ \ ({{\rm log}\al})^2
\end{equation}
with some appropriate integration constants. Notice that the behavior 
$t \sim c_1{\rm log}(c_2\al)$ coincides with the 
desired behavior of the mirror map
at large radius limit.
So, instead 
of solving them directly, we assume that 
the solution $t = t(\al)$ is given by the mirror map whose
explicit form is known to be
\begin{eqnarray}
t(\al) &=&
{2(\bar{w}_1^{(0)} - \bar{w}_0^{(0)}) + \bar{w}_2^{(0)} -
\bar{w}_4^{(0)}\over 5 \bar{w}_0^{(0)}}, \cr
t(\al) &=&
{\bar{w}_2^{(0)} - \bar{w}_0^{(0)} + \bar{w}_1^{(0)} - \bar{w}_5^{(0)}
\over 3 \bar{w}_0^{(0)}}, \cr
t(\al) &=&
{\bar{w}_2^{(0)} - \bar{w}_0^{(0)} + \bar{w}_1^{(0)} - \bar{w}_7^{(0)}
\over 2 \bar{w}_0^{(0)}},\cr
t(\al) &=&
{\bar{w}_2^{(0)} - \bar{w}_9^{(0)}
\over \bar{w}_0^{(0)}}
\label{stasta}
\end{eqnarray}
for $k = 5,6,8$ and $10$, respectively \cite{KT}.
Substituting these mirror maps and $r_{11} = {\bar w}_0^{(0)}$
into (\ref{def}) we
have checked explicitly that the mirror map (\ref{stasta})
satisfies (\ref{def}). Thus we observe that the 
differential equations for the periods and the mirror map
are obtained from the Gauss-Manin system in the flat coordinates.

The mirror map (\ref{stasta}) determines the Yukawa coupling (\ref{yc})
up to an arbitrary constant $C$. This constant is fixed by 
the well-known fact that the Yukawa coupling in the
large radius limit is given by the intersection numbers of a basis 
for $H^2(M_{\al},{\bf Z})$.

Finally, we present the prepotential within this 
framework. The Gauss-Manin system (\ref{fcgm}) 
in the flat coordinates 
is equivalent to
\begin{equation}
{\del^2 \over \del t^2}
\left(
{1 \over K_{ttt}(t)} 
{\del^2 \over \del t^2} v^{(0)}
\right) = 0.
\label{rgm}
\end{equation}
For $v^{(0)} = {}^t(v_0^{(0)},v_1^{(0)},v_2^{(0)},v_3^{(0)})$,
eq.(\ref{rgm}) reduces to 
\begin{equation}
v_0^{(0)} = 1, \ v_1^{(0)} = t, \ 
{\del^2 \over \del t^2} v_2^{(0)} = K_{ttt}, \
{\del^2 \over \del t^2} v_3^{(0)} = - t K_{ttt}.
\end{equation}
It is then straightforward to see that the prepotential is
constructed as
\begin{equation}
F(t) = {1 \over 2}(
v_0^{(0)} v_3^{(0)} + v_1^{(0)} v_2^{(0)} )
\end{equation}
such that
$K_{ttt}(t) = {\del^3 \over \del t^3} F(t)$ \cite{HL}.
{}From the transformation matrix $M_k$ (\ref{tmat}), we know the relation 
$v_{\gamma_j}^{(0)} = w_{\gamma_j}^{(0)}/r_{11}
=w_{\gamma_j}^{(0)}/{\bar w}_0^{(0)}$.
Therefore we can determine the prepotential from the solutions of
the Picard-Fuchs equation.

\setcounter{equation}{0}
\section{Flat coordinates for a two-parameter model}

In this section, we consider a two-parameter model. 
A mirror symmetry of two-parameter models have been intensively studied
by Hosono et al.$\ $\cite{HKTY} 
and by Candelas et al.$\ $\cite{COFKM} \cite{CFKM}.
In the following we concentrate on the model described by
\begin{equation}
M = \{x_i \in {\bf P}(1,1,2,2,2)|W = x_0^8 + x_1^8 + x_2^4 + x_3^4 + x_4^4 \}.
\label{2para}
\end{equation}
This Calabi-Yau manifold has 
the Hodge numbers $h_{1,1} = 2$ and $h_{2,1} = 86$.

To obtain the mirror manifold of this model, one must consider the
orbifoldization of $M$ by dividing out a discrete symmetry $G = {\bf Z}_4^3$.
The generators of $G$ are given by
\begin{eqnarray}
{\bf Z}_4&:& \ (x_0,x_1,x_2,x_3,x_4) \rightarrow 
             (x_0,e^{{2\pi i \over 4}3}x_1,e^{2\pi i \over 4}x_2,x_3,x_4)\cr
{\bf Z}_4&:& \ (x_0,x_1,x_2,x_3,x_4) \rightarrow 
             (x_0,e^{{2\pi i \over 4}3}x_1,x_2,e^{2\pi i \over 4}x_3,x_4)\cr
{\bf Z}_4&:& \ (x_0,x_1,x_2,x_3,x_4) \rightarrow 
             (x_0,e^{{2\pi i \over 4}3}x_1,x_2,x_3,e^{2\pi i \over 4}x_4).
\end{eqnarray}
There are two marginal operators (degree eight monomials)
which are invariant
under this discrete symmetry. So, one can deform the defining relation
in (\ref{2para}) with these operators. This deformation is regarded as the
complex structure deformation of the mirror manifold of (\ref{2para}).
Now the deformed polynomial is 
given by
\begin{equation}
W(\al,\be) = W - 8\al x_0 x_1 x_2 x_3 x_4 - 2 \be x_0^4 x_1^4,
\end{equation}
where $\al$ and $\be$ are deformation parameters.
Let $M_{\al,\be}$ denote the manifold which is defined by 
the equation $W(\al,\be) = 0$ on ${\bf P}(1,1,2,2,2)$, then
one finds that the mirror manifold of $M_{\al,\be}$ is realized as
\begin{equation}
\bar{M}_{\al,\be} = M_{\al,\be}/G.
\end{equation}
This manifold actually has the Hodge numbers
$h_{1,1}=86$ and $h_{2,1} = 2$.

We introduce the deformed Jacobian ring like (\ref{djr})
to discuss the quantum cohomology ring of $\bar {M}_{\al,\be}$
using the Gauss-Manin system.
As the basis of the ring we choose
\begin{eqnarray}
&1,&\cr
&x_0x_1x_2x_3x_4, \ \ x_0^4x_1^4,& \cr
&x_0^2x_1^2x_2^2x_3^2x_4^2, \ \ x_0^5 x_1^5 x_2 x_3 x_4,& \cr
&x_0^6 x_1^6 x_2^2 x_3^2 x_4^2.&
\end{eqnarray}
The period integrals are then defined by
\[
w_{\gamma_j}^{(0)} = 
\int_{\gamma_j} {1\over W(\al,\be)}d\mu,
\]
\[
w_{\gamma_j}^{(1)} = 
- \int_{\gamma_j} {x_0 x_1 x_2 x_3 x_4\over W^2(\al,\be)}d\mu, \ \
w_{\gamma_j}^{(2)} = 
- \int_{\gamma_j} {x_0^4 x_1^4 \over W^2(\al,\be)}d\mu,
\]
\[
w_{\gamma_j}^{(3)} = 
2 \int_{\gamma_j} {x_0^2 x_1^2 x_2^2 x_3^2 x_4^2 \over W^3(\al,\be)}d\mu, \ \
w_{\gamma_j}^{(4)} = 
2 \int_{\gamma_j} {x_0^5 x_1^5 x_2 x_3 x_4 \over W^3(\al,\be)}d\mu,
\]
\begin{equation}
w_{\gamma_j}^{(5)} = 
-6 \int_{\gamma_j} {x_0^6 x_1^6 x_2^2 x_3^2 x_4^2 \over W^4(\al,\be)}d\mu,
\label{2parapi}
\end{equation}
where $\gamma_0,\gamma_1,\cdots,\gamma_5$ are certain
homology cycles in $H_3({\bar M}_{\al,\be},{\bf Z})$.
For the two-parameter model,
these period integrals satisfy two sets
of the first order differential equations
\begin{equation}
{\del \over \del \al}w = G(\al,\be)w, \ \
{\del \over \del \be}w = H(\al,\be)w,
\label{gm2para}
\end{equation}
where $w={}^t(w^{(0)},w^{(1)},w^{(2)},w^{(3)},w^{(4)},w^{(5)})$.
Making use of  the relations obtained from 
$\del W(\al,\be)/\del x_i = \dots$ and 
integrating by parts, we determine the matrices 
$G$ and $H$ 
\begin{eqnarray}
G(\al,\be) &=&
\left(
\begin{array}{cccccc}
0&-8&0&0&0&0 \cr
0&0&0&-8&0&0 \cr
0&0&0&0&-8&0 \cr
0&0&-\al&0&24\al^2&-64\al^3 \cr
0&0&0&0&0&-8 \cr
{\al\over 64 \Delta}&{-15\al^2\over 8\Delta}&
{-\al(44\al^4+3\be)\over 4\Delta}&{25\al^3 \over \Delta}&
{2\al^2(148\al^4+13\be)\over\Delta}&{-128\al^3(8\al^4+\be)\over \Delta}
\end{array}
\right), \cr
H(\al,\be) &=&
\left(
\begin{array}{cccccc}
0&0&-2&0&0&0 \cr
0&0&0&0&-2&0 \cr
{-1\over 32 Z}&{3\al\over 4Z}&{3 \be \over 2Z}&{- 2\al^2 \over Z}&
{-4\al\be\over Z}&0 \cr
0&0&0&0&0&-2 \cr
0&{-1\over 8Z}&{-\al^3\over 4Z}&{5\al\over 4Z}&{2(3\al^4+\be)\over Z}&
{-4\al (4\al^4+\be)\over Z}\cr
{\al^2 (4\al^4 + \be)\over 128 Z \Delta}&
{-15\al^3(4\al^4+\be)\over 16Z\Delta}&
H_{63}&H_{64}&H_{65}&H_{66}
\end{array}
\right),
\label{GH}
\end{eqnarray}
where
\begin{eqnarray}
\Delta &=& (8 \al^4 + \be)^2 -1, \ Z = 1 - \be^2, \cr
H_{63} &=& {-\al^2 (48\al^8 + 24 \al^4 \be + \be^2 +2)\over 8Z\Delta},\cr
H_{64} &=& {9+1024\al^8 + 256 \al^4 \be - 9\be^2\over 32Z\Delta},\cr
H_{65} &=& {\al^3 (7+144\al^8+88\al^4\be + 6\be^2)\over Z\Delta},\cr
H_{66} &=&{-52\al^4 - 768\al^{12} - 5\be - 384\al^8 \be 
+ 4\al^4 \be^2 + 5\be^3
\over 2Z\Delta}.
\end{eqnarray}
These expressions have been obtained in \cite{COFKM}.

{}From eqs.(\ref{gm2para}) and 
(\ref{GH}), one obtains five differential equations
satisfied by the period $w^{(0)}$.
However, some of them are not independent. 
By careful analysis, we find a set of independent equations.
These equations are the Picard-Fuchs equations for $w^{(0)}$
whose explicit forms are
\begin{eqnarray}
& &
{w^{(0)}}^{(3,0)} - 32\al^3 {w^{(0)}}^{(2,1)} - 96\al^2 {w^{(0)}}^{(1,1)} - 
32\al {w^{(0)}}^{(0,1)} = 0, \cr
& & 16(\be^2 - 1) {w^{(0)}}^{(0,2)} 
+ \al^2 {w^{(0)}}^{(2,0)} + 8 \al\be {w^{(0)}}^{(1,1)} +
3 \al {w^{(0)}}^{(1,0)} + 24 \be {w^{(0)}}^{(0,1)} + {w^{(0)}} = 0, \cr
& &(8 \al^4 + \be + 1)(8\al^4 + \be -1) {w^{(0)}}^{(3.1)} + 
128\al^3 (8\al^4 + \be) {w^{(0)}}^{(2,1)} \cr
& &
\ \ \ 
+ 50\al^3 {w^{(0)}}^{(2,0)} + 16\al^2(148\al^4 + 13\be){w^{(0)}}^{(1,1)} +
30\al^2 {w^{(0)}}^{(1,0)} \cr
& &
\ \ \ + 16\al(44\al^4 + 3\be){w^{(0)}}^{(0,1)} + 2\al {w^{(0)}} = 0,
\label{pf2para}
\end{eqnarray}
where ${w^{(0)}}^{(i,j)} = {\del^{i+j} {w^{(0)}} \over \del \al^i \del \be^j}$ 
\cite{COFKM}. 

Now, to characterize the flat coordinates, we make a degree-preserving 
change of our basis of the Jacobian ring as
\begin{equation}
\begin{array}{c}
1,\cr
x_0x_1x_2x_3x_4, \ \ x_0^4x_1^4, \cr
x_0^2x_1^2x_2^2x_3^2x_4^2, \ \ x_0^5 x_1^5 x_2 x_3 x_4, \cr
x_0^6 x_1^6 x_2^2 x_3^2 x_4^2,
\end{array}
\rightarrow
\begin{array}{c}
{\cal O}^{(0)},\cr
{\cal O}_t^{(1)}, \ \ {\cal O}_s^{(1)}, \cr
{\cal O}^{s(2)}, \ \ {\cal O}^{t(2)}, \cr
{\cal O}^{(3)},
\end{array}
\end{equation}
with ${\cal O}^{(0)} = 1$. 
The superscript
stands for the charge of ${\cal O}^{(i)}$ 
which is a polynomial of degree $8i$.
Correspondingly we regard $\al$ and $\be$ as functions of $t$ and $s$,
i.e. $\al = \al(t,s)$ and  $\be = \be(t,s)$.
For this new basis, we can define the period integrals as given by 
(\ref{fcpi}). 
Let $v$ denote the period vector in this coordinate, then
$v$ is related to $w$ 
through
\begin{equation}
w(\al,\be)  =  M(\al,\be)v(t,s),
\label{tm2para}
\end{equation}
where
\begin{equation}
M(\al,\be)=
\left(
\begin{array}{cccccc}
r_{11}&0&0&0&0&0 \cr
r_{21}&r_{22}&r_{23}&0&0&0 \cr
r_{31}&r_{32}&r_{33}&0&0&0 \cr
r_{41}&r_{42}&r_{43}&r_{44}&r_{45}&0 \cr
r_{51}&r_{52}&r_{53}&r_{54}&r_{55}&0 \cr
r_{61}&r_{62}&r_{63}&r_{64}&r_{65}&r_{66}
\end{array}
\right).
\end{equation}
Notice that since there are two elements in charge one and two sectors,
respectively,
the matrix elements $r_{23}$ and $r_{45}$ are allowed to exist.

We constrain the basis $\{ {\cal O}^{(0)},{\cal O}_t^{(1)},
{\cal O}_s^{(1)},{\cal O}^{s(2)},{\cal O}^{t(2)},{\cal O}^{(3)}\}$
to satisfy the flat coordinate conditions
\begin{eqnarray}
{\rm i)}& &{\cal O}_t^{(1)} = {\del \over \del t}(r_{11}W),
\ \ \
         {\cal O}_s^{(1)} = {\del \over \del s}(r_{11}W), \cr
{\rm ii)}& &{\cal O}_i^{(1)} {\cal O}_j^{(1)} = K_{ijk} {\cal O}^{k(2)}, \
         (i,j,k = s,t), \cr
{\rm iii)}& &{\cal O}_i^{(1)} {\cal O}^{j(2)} = \delta_i^j {\cal O}^{(3)}, \
          (i,j = s,t),\cr
{\rm iv)}& &{\cal O} {\cal O}^{(3)} = 0, \ 
({\cal O} = {\cal O}^{(1)}, {\cal O}^{(2)}).
\end{eqnarray}
{}From these conditions, 
we derive two first order differential equations for $v$
with respect to $t$ and $s$. The result is
\begin{equation}
{\del \over \del t} v = R_t v, \ \ \  {\del \over \del s} v = R_s v,
\label{fcgmst}
\end{equation}
where
\begin{equation}
R_t
=
  \left(
  \begin{array}{cccccc}
  0&1&0&0&0&0 \cr
  0&0&0&K_{tts}&K_{ttt}&0 \cr
  0&0&0&K_{tss}&K_{tst}&0 \cr
  0&0&0&0&0&0 \cr
  0&0&0&0&0&1 \cr
  0&0&0&0&0&0
  \end{array}
  \right) 
\label{fcgmt}
\end{equation}
and 
\begin{equation}
R_s
=
  \left(
  \begin{array}{cccccc}
  0&0&1&0&0&0 \cr
  0&0&0&K_{sts}&K_{stt}&0 \cr
  0&0&0&K_{sss}&K_{sst}&0 \cr
  0&0&0&0&0&1 \cr
  0&0&0&0&0&0 \cr
  0&0&0&0&0&0
  \end{array}
  \right).
\label{fcgms}
\end{equation}

Using (\ref{tm2para}), (\ref{fcgmst}), $R_t$ and $R_s$, 
the original Gauss-Manin system (\ref{gm2para}) is rewritten as
\begin{equation}
A v = 0, \ \ \  B v = 0,
\label{AB}
\end{equation}
where
\begin{equation}
A = \left( G M - {\del M \over \del \al} - M {\del t \over \del \al} R_t - M
{\del s \over \del \al} R_s \right) 
\label{fcgm1}
\end{equation}
and
\begin{equation}
B = \left( H M - {\del M \over \del \be} - M {\del t \over \del \be} R_t - M
{\del s \over \del \be} R_s \right).
\label{fcgm2}
\end{equation}
Since all the elements of $v$ are linearly independent eq.(\ref{AB})
is equivalent to 
\begin{equation}
A(r_{11},t,s)=0, \ \ \ B(r_{11},t,s)=0
\end{equation}
which give rise to the differential equations obeyed by 
$r_{11}(\al,\be)$, $t(\al,\be)$ and $s(\al,\be)$.

To write down them explicitly we first determine the 
transformation matrix $M(\al,\be)$.
From eqs.(\ref{AB})$\sim$(\ref{fcgm2}), we get
\begin{equation}
\begin{array}{l}
M(\al,\be) = \cr
{}\cr
\left(
\begin{array}{cccccc}
r_{11}&0&0&0&0&0 \cr
-{1\over 8}{\del \over \del \al}r_{11}&
-{1\over 8}r_{11} {\del \over \del \al}t&
-{1\over 8}r_{11} {\del \over \del \al}s&0&0&0 \cr
-{1\over 2}{\del \over \del \be}r_{11}&
-{1\over 2}r_{11} {\del \over \del \be}t&
-{1\over 2}r_{11} {\del \over \del \be}s&0&0&0 \cr
{1\over 64}{\del^2 \over \del \al^2}r_{11}&
{1\over 64}r_{11} {\del^2 \over \del \al^2}t +
{1\over 32}{\del \over \del \al}r_{11} {\del \over \del \al}t
&
{1\over 64}r_{11} {\del^2 \over \del \al^2}s +
{1\over 32}{\del \over \del \al}r_{11} {\del \over \del \al}s
&r_{44}&r_{45}&0 \cr
{1\over 16}{\del^2 \over \del \al \del \be} r_{11}&
r_{52}&r_{53}&r_{54}&r_{55}&0 \cr
-{1\over 128}{\del^3 \over \del \al^2 \del \be}r_{11}&
r_{62}&r_{63}&r_{64}&r_{65}&r_{66}
\end{array}
\right),
\end{array}
\label{tm2}
\end{equation}
where the explicit forms of $r_{ij}$ are given in appendix
(see eqs.(\ref{app1})$\sim$(\ref{app2})).
Substituting this transformation matrix into (\ref{fcgm1}) and (\ref{fcgm2}),
we see $K_{tts}=K_{tst}=K_{stt}$ and $K_{sst}=K_{sts}=K_{tss}$.
If we define
\[
K_{\al\al\al} = - {4096 \al^3 \over \Delta}, \ \
K_{\al\al\be} = - {128 \over \Delta},
\]
\begin{equation}
K_{\al\be\be} = - {64\al(\be + 4\al^4) \over (1 - \be^2) \Delta}, \ \
K_{\be\be\be} 
= - {8\al^2 (1 + 3\be^2 + 16 \al^4 \be) \over (1 - \be^2)^2 \Delta},
\end{equation}
where $\Delta = (8 \al^4 + \be)^2 - 1$, 
we find 
\[
K_{ttt} = {C \over r_{11}^2}
\left[
K_{\al\al\al}\left({\del \al \over \del t}\right)^3 +
3 K_{\al\al\be}\left({\del \al \over \del t}\right)^2 {\del \be \over \del t}+
3 K_{\al\be\be}{\del \al \over \del t}\left({\del \be \over \del t}\right)^2 +
K_{\be\be\be}\left({\del \be \over \del t}\right)^3
\right],
\]
\begin{eqnarray}
K_{tts}&=&
{C\over r_{11}^2}
\left[
K_{\al\al\al}\left({\del \al \over \del t}\right)^2 {\del \al\over  \del s} +
K_{\al\al\be}\left(
\left( {\del \al \over \del t}\right)^2 {\del \be \over \del s} +
2 {\del \al \over \del t}{\del \al \over \del s}
{\del \be \over \del t}\right)\right. \cr
& &\left.
+ K_{\al\be\be}\left(
{\del \al \over \del s}\left({\del \be \over \del t}\right)^2 +
2 {\del \al \over \del t}{\del \be \over \del s}
{\del \be \over \del t}\right) +
K_{\be\be\be}\left({\del \be \over \del t}\right)^2{\del \be \over \del s}
\right],
\label{yc2para}
\end{eqnarray}
where $C$ is an arbitrary constant and 
we have used
\begin{equation}
\left(
\begin{array}{cc}
{\del \al \over \del t} & {\del \be \over \del t} \cr
{\del \al \over \del s} & {\del \be \over \del s}
\end{array}
\right)
=
\left(
\begin{array}{cc}
{\del t \over \del \al} & {\del s \over \del \al} \cr
{\del t \over \del \be} & {\del s \over \del \be}
\end{array}
\right)^{-1}
=
{1\over {\del t \over \del \al}{\del s \over \del \be}
       -{\del t \over \del \be}{\del s \over \del \al}}
\left(
\begin{array}{rr}
{\del s \over \del \be}& -{\del s \over \del \al} \cr
-{\del t \over \del \be} & {\del t \over \del \al}
\end{array}
\right) 
\end{equation}
to convert 
${\del t \over \del \al},{\del t \over \del \be},
 {\del s \over \del \al},{\del s \over \del \be}$ into 
${\del \al \over \del t},{\del \al \over \del s},
 {\del \be \over \del t},{\del \be \over \del s}$.
$K_{sss}$ and $K_{tss}$ are obtained from $K_{ttt}$ and $K_{tts}$
by interchanging $t$ and $s$.
The form of (\ref{yc2para}) is in agreement with \cite{COFKM}.
Substituting (\ref{tm2}) and $K_{ijk}$ 
into (\ref{fcgm1}) and (\ref{fcgm2}), we find 
the matrices $A$ and $B$ as follows
\begin{equation}
A =
\left(
\begin{array}{cccccc}
0&0&0&0&0&0 \cr
0&0&0&0&0&0 \cr
0&0&0&0&0&0 \cr
A_{41}(r_{11})&A_{42}(r_{11},t)&A_{43}(r_{11},s)&A_{44}(r_{11},t,s)&
A_{45}(r_{11},t,s)&0 \cr
0&0&0&A_{54}(r_{11},t,s)&A_{55}(r_{11},t,s)&0 \cr
A_{61}(r_{11})&A_{62}(r_{11},t)&A_{63}(r_{11},s)&A_{64}(r_{11},t,s)&
A_{65}(r_{11},t,s)&0 
\end{array}
\right)
\end{equation}
and
\begin{equation}
B =
\left(
\begin{array}{cccccc}
0&0&0&0&0&0 \cr
0&0&0&0&0&0 \cr
B_{31}(r_{11})&B_{32}(r_{11},t)&B_{33}(r_{11},s)&0&0&0 \cr
0&0&0&B_{44}(r_{11},t,s)&
B_{45}(r_{11},t,s)&0 \cr
B_{51}(r_{11})&B_{52}(r_{11},t)&B_{53}(r_{11},s)
&B_{54}(r_{11},t,s)&B_{55}(r_{11},t,s)&0 \cr
B_{61}(r_{11})&B_{62}(r_{11},t)&B_{63}(r_{11},s)&B_{64}(r_{11},t,s)&
B_{65}(r_{11},t,s)&0
\end{array}
\right).
\end{equation}
Now, $A_{ij}=0$ and $B_{ij}=0$ for all $i,j$ 
are non-trivial differential equations which
must be satisfied by $t,s$ and $r_{11}$. 
Explicit forms of $A_{ij}$ and $B_{ij}$ are given in appendix.

Here we explain some detailed properties of $A_{ij}$ and $B_{ij}$. 

(i) The relations for the elements of the first column:
The equations
$A_{41}(r_{11})=0$, $A_{61}(r_{11})=0$ and $B_{31}(r_{11})=0$ are
proportional to the Picard-Fuchs equations (\ref{pf2para})
(see eqs.(\ref{app3})$\sim$(\ref{app4}) in appendix).
So, $r_{11}$ is a solution of the Picard-Fuchs equations.
We also notice that 
$B_{51}(r_{11})$ and $B_{61}(r_{11})$ are not independent.
They are expressed in terms of 
$A_{41}(r_{11})$, $A_{61}(r_{11})$, $B_{31}(r_{11})$
and their derivatives.

(ii) The relations for the elements of the second and third columns:
We have
\begin{equation}
A_{i2}(r_{11},t) = A_{i3}(r_{11},t)
\label{ii1}
\end{equation}
for $i=4,6$ and
\begin{equation}
B_{i2}(r_{11},t) = B_{i3}(r_{11},t)
\label{ii2}
\end{equation}
for $i=3,5$ and $6$.
In addition, $B_{52}(r_{11},t)$ and $B_{62}(r_{11},t)$ are not independent.
They are expressed in terms of 
$A_{42}(r_{11},t)$, $A_{62}(r_{11},t)$, $B_{32}(r_{11},t)$ and
their derivatives.
Moreover, let ${\bar w}_j^{(0)}\ (0 \leq j \leq 5)$ denote the solutions of 
the Picard-Fuchs equations (\ref{pf2para}), then we find that 
the solutions of 
$A_{42}(r_{11},t) = 0$, $A_{62}(r_{11},t)=0$ and $B_{32}(r_{11},t) =0$
are expressed by 
\begin{equation}
t = \sum_{j=0}^5 {c_j {\bar w}_j^{(0)} \over r_{11}},
\label{sol1}
\end{equation}
where $c_j$'s are arbitrary constants. 
Similarly, for $s$, we obtain
\begin{equation}
s = \sum_{j=0}^5 {a_j {\bar w}_j^{(0)} \over r_{11}}
\label{sol2}
\end{equation}
with $a_j$ being arbitrary constants.

(iii) The relations for the elements of the fourth and fifth columns:
We have
\begin{equation}
A_{i4}(r_{11},t,s) = A_{i5}(r_{11},s,t), \ \ \ 
B_{i4}(r_{11},t,s) = B_{i5}(r_{11},s,t)
\end{equation}
for $i = 4,5$ and $6$. We also find
\begin{equation}
4 {\del t \over \del \be} A_{44}(r_{11},t,s)=
   {\del t \over \del \al} A_{54}(r_{11},t,s).
\end{equation}
Hence, among the elements of the fourth and fifth columns, 
we are left with five elements, say $A_{44}$, $A_{64}$, $B_{44}$, $B_{54}$
and $B_{64}$, which seem to be independent. Are these five elements
really independent? As long as we have manipulated these equations
explicitly we feel that there still exist some relations which can be
used to reduce the number of independent equations.
Unfortunately the equations we have to deal with
are so complicated that we have not yet succeeded in finding them.

To summarize, analyzing the Gauss-Manin system in the flat coordinates 
we have obtained the equations for $r_{11}$
\begin{equation}
A_{41}(r_{11}) = 0,\ \ \
A_{61}(r_{11}) = 0,\ \ \
B_{31}(r_{11}) = 0,
\label{*1}
\end{equation}
which are equivalent to the Picard-Fuchs equations (\ref{pf2para}).
This is the property (i).
The property (ii) is that the flat coordinates $t$ and $s$ take the form
(\ref{sol1}) and (\ref{sol2}), respectively.
In order to fix $c_j$ and $a_j$
we have to solve
\begin{eqnarray}
&A_{44}(r_{11},t,s)=0,\ \ \
A_{64}(r_{11},t,s)=0,\ \ \
B_{44}(r_{11},t,s)=0,& \cr
&B_{54}(r_{11},t,s)=0,\ \ \ 
B_{64}(r_{11},t,s)=0&
\label{*2}
\end{eqnarray}
according to the property (iii),
though we are not sure if these five equations are all independent.

We now wish to solve our differential equations (\ref{*1}) and (\ref{*2}).
First of all, from the explicit calculations of period integral and 
the analytic continuation, one can find a solution 
to the Picard-Fuchs equations. The result is \cite{COFKM} \cite{BCOFHJQ}
\begin{equation}
\bar{w}^{(0)}_0(\al,\be) = - {1\over 4\al}
\sum^\infty_{n=1}{\Gamma({n\over 4})(-8\al)^n 
\over \Gamma(n)\Gamma^3 (1-{n\over 4})}U_{-{n\over 4}}(\be),\ \  |\al| \ll 1
\end{equation}
where
\begin{equation}
U_{\nu}(\be) = {e^{i\pi\nu/2} \over 2 \Gamma(-\nu)}
\sum^\infty_{m=0}{e^{i\pi m/2}\Gamma({m-\nu \over 2})
\over m! \Gamma(1 - {m-\nu \over 2})}(2\be)^m,\ \  |\be| < 1.
\end{equation}
Then the other solutions are given by
\begin{equation}
\bar{w}^{(0)}_j(\al,\be) = \bar{w}^{(0)}_0(e^{{i\pi\over 4}j}\al,(-1)^j\be)
\end{equation}
for $j=1,2,\cdots,5$.

Let us next turn to (\ref{*2}). We notice that the forms of 
$t$ and $s$ in (\ref{sol1}), (\ref{sol2}) look like the mirror map for 
the two-parameter model if we take $r_{11} \propto {\bar w}_0^{(0)}$. 
In fact the mirror map found by Candelas et al. is given by \cite{COFKM}
\begin{equation}
t(\al,\be) = -{1\over 2} +{1\over 4 \bar{w}^{(0)}_0}
[(\bar{w}^{(0)}_0 + 2\bar{w}^{(0)}_2 + \bar{w}^{(0)}_4) + 
 (3\bar{w}^{(0)}_1 + 2\bar{w}^{(0)}_3 + \bar{w}^{(0)}_5)],
\end{equation}
\begin{equation}
s(\al,\be) = -{1\over 2} +{1\over 4 \bar{w}^{(0)}_0}
[(\bar{w}^{(0)}_0 + 2\bar{w}^{(0)}_2 + \bar{w}^{(0)}_4) -
 (3\bar{w}^{(0)}_1 + 2\bar{w}^{(0)}_3 + \bar{w}^{(0)}_5)].
\end{equation}
Thus we assume
\begin{equation}
r_{11} = {\bar w}_0^{(0)}(\al,\be), \ \ \
t = t(\al,\be), \ \ \
s = s(\al,\be) 
\label{*3}
\end{equation}
and check if these satisfy (\ref{*2}). Actually,
after tedious calculations, we have confirmed that (\ref{*2}) is
satisfied by (\ref{*3}).
This is an extremely non-trivial result.
Thus, in the two-parameter case, we have also derived the differential 
equations which govern the periods and the mirror maps. 

It should be remarked that
$t = t(\al,\be)$ and $s = s(\al,\be)$ are not unique 
solutions of (\ref{*2}). In fact 
${\tilde t} = a \ t(\al,\be) + b \ s(\al,\be)$
and ${\tilde s} = c \ t(\al,\be) + d \ s(\al,\be)$ 
are also solutions, where
$a,b,c$ and $d$ are arbitrary constants with $a d \neq b c$. 
We discuss this point at the end of appendix.

Finally we derive the prepotential from 
the solutions of
the Gauss-Manin system 
in the flat coordinates (\ref{fcgmst}). 
It is not difficult to see that
eq.(\ref{fcgmst}) is equivalent to
\begin{eqnarray}
{\del^2 \over \del x \del y}
\left\{
{K_{sss} \over K_{ttt} K_{sss} - K_{tts} K_{sst}}
{\del^2 \over \del t^2} -
{K_{tts} \over K_{ttt} K_{sss} - K_{tts} K_{sst}}
{\del^2 \over \del s^2}
\right\} v^{(0)} &=& 0,\cr
{\del^2 \over \del x \del y}
\left\{
{K_{tss} \over K_{ttt} K_{tss} - K_{tts}^2}
{\del^2 \over \del t^2} -
{K_{tts} \over K_{ttt} K_{tss} - K_{tts}^2}
{\del^2 \over \del s \del t}
\right\} v^{(0)} &=& 0,\cr
{\del^2 \over \del x \del y}
\left\{
{K_{tts} \over K_{ttt} K_{tss} - K_{tts}^2}
{\del^2 \over \del t^2} -
{K_{ttt} \over K_{ttt} K_{tss} - K_{tts}^2}
{\del^2 \over \del s \del t}
\right\} v^{(0)} &=& 0
\label{tuka}
\end{eqnarray}
with $x,y = s,t$, and those obtained from
(\ref{tuka}) by interchanging $t$ and $s$.
For $v^{(0)} = 
{}^t(v_0^{(0)},v_1^{(0)},v_2^{(0)},v_3^{(0)},v_4^{(0)},v_5^{(0)})$,
eq.(\ref{tuka}) reduces to
\begin{eqnarray}
&v_0^{(0)} = 1,\ v_1^{(0)} = t, \ v_2^{(0)} = s,&\cr
v_3^{(0)}:& \ {\del^2 \over \del t^2} v_3^{(0)} = K_{stt}, \
              {\del^2 \over \del s \del t} v_3^{(0)} = K_{tss}, \
              {\del^2 \over \del s^2} v_3^{(0)} = K_{sss},& \cr 
v_4^{(0)}:& \ {\del^2 \over \del t^2} v_4^{(0)} = K_{ttt}, \
              {\del^2 \over \del s \del t} v_4^{(0)} = K_{tts}, \
              {\del^2 \over \del s^2} v_4^{(0)} = K_{tss},& \cr
v_5^{(0)}:&
 {\del^2 \over \del t^2} v_5^{(0)} = -(t K_{ttt} + s K_{stt}),& \cr
&{\del^2 \over \del s \del t} v_5^{(0)} = -( t K_{tts} + s K_{sst}),& \cr
&{\del^2 \over \del s^2} v_5^{(0)} = -(t K_{tss} + s K_{sss}).&
\end{eqnarray}
Then it is easy to show that the prepotential is given by
\begin{equation}
F(t,s) = {1\over 2} ( v_0^{(0)} v_5^{(0)} + v_1^{(0)} v_4^{(0)} +
                      v_2^{(0)} v_3^{(0)}).
\end{equation}
Of course, this is obtained so as to have
$K_{xyz} = {\del^3 \over \del x \del y \del z} F(t,s)$, where $x,y,z = s,t$.
From the transformation matrix (\ref{tm2}), we know the relation 
$v_{\gamma_j}^{(0)} =  w_{\gamma_j}^{(0)}/r_{11} 
= w_{\gamma_j}^{(0)}/{\bar w}_0^{(0)}$,
from which we can evaluate the prepotential.

\setcounter{equation}{0}

\section{Conclusions}

In this article we have studied mirror symmetry of Calabi-Yau
manifolds by formulating the 
Gauss-Manin system in the flat coordinates. 
Following the work by Hosono-Lian \cite{HL} on the quintic hypersurface
we have shown for various examples of Calabi-Yau manifolds
that the Gauss-Manin system yields the differential equations obeyed by
the mirror map in addition to the ordinary Picard-Fuchs equations.

It is very interesting that the mirror map is
governed by differential equations such as (\ref{def})
for one-parameter models
and (\ref{*2}) for a two-parameter model.
Concerning the two-parameter model we expect that the system
of equations (\ref{*2}) is still redundant and  it
may reduce to more fundamental one.

When checking (\ref{def}) and (\ref{*2}) we have assumed the 
form of the mirror map.
It is desirable to find a way to solve (\ref{def}) and (\ref{*2})
so that one can construct the mirror map directly as a solution to
the differential equations. Furthermore 
it will be important to understand the deeper structure and the physical 
meaning of the differential equations for the mirror map in view
of duality symmetry.
\newline

\section*{Acknowledgements}

I would like to thank S. Hosono for 
excellent lectures on Calabi-Yau manifolds and
mirror symmetry. I also wish to thank 
S.-K. Yang for useful discussions, suggestions and 
encouragements.
\newline

\appendix

\renewcommand{\theequation}{A.\arabic{equation}}
\setcounter{equation}{0}
\section*{Appendix}

In this appendix, we present the matrix elements of $M(\al,\be)$, 
$A$ and $B$ for the two-parameter model in section 3.
To simplify the expressions, we introduce the following notations
\begin{equation}
Z = 1 -\be^2, \ \Delta = 64\al^8 + 16 \al^4 \be + \be^2 - 1, \ D =
{\del t \over \del \al}  {\del s\over \del \be} - {\del t\over \del \be}
{\del s\over \del \al}.
\end{equation}

First, we give the elements of $M(\al,\be)$.
They are 
\begin{equation}
r_{44} = 
{2 C \over \Delta \ D \ r_{11}}
\left(
32 \al^3 {\del t\over \del \be } - {\del t \over \del \al}
\right),
\label{app1}
\end{equation}
\begin{equation}
r_{45} = 
{- 2 C \over \Delta \ D \ r_{11}}
\left(
32 \al^3 {\del s\over \del \be } - {\del s\over \del \al}
\right)
,
\end{equation}
\begin{equation}
r_{52} =
{1 \over 16}r_{11}{\del^2 t\over \del \al \del \be} +
{1 \over 16} {\del r_{11}\over \del \al} {\del t\over \del \be} +
{1 \over 16} {\del r_{11}\over \del \be} {\del t\over \del \al},
\end{equation}
\begin{equation}
r_{53} =
{1 \over 16}r_{11}{\del^2 s\over \del \al \del \be} +
{1 \over 16} {\del r_{11}\over \del \al} {\del s\over \del \be} +
{1 \over 16} {\del r_{11}\over \del \be} {\del s\over \del \al},
\end{equation}
\begin{equation}
r_{54}=
{- 4 C \over Z \ \Delta \ D \ r_{11}}
\left(
4 \al^5 {\del t \over \del \al} + \al \be {\del t \over \del \al} 
- 2 {\del t \over \del \be} + 2 \be^2 {\del t\over \del \be}
\right)
,
\end{equation}
\begin{equation}
r_{55}=
{ 4 C \over Z \ \Delta \ D \ r_{11}}
\left(
4 \al^5 {\del s \over \del \al}  + \al \be {\del s \over \del \al}
- 2 {\del s \over \del \be} + 2 \be^2 {\del s\over \del \be}
\right)
,
\end{equation}
\begin{eqnarray}
r_{62} &=&
-{1\over 128} r_{11} {\del^3 t\over \del \al^2 \del \be}  -
{1 \over 64} {\del r_{11}\over \del \al} {\del^2 t\over \del \al \del \be}-
{1 \over 128} {\del r_{11}\over \del \be}{\del^2 t\over \del \al^2}  \cr
& &-{1 \over 128} {\del^2 r_{11}\over \del \al^2}{\del t \over \del \be} -
{1 \over 64} {\del^2 r_{11}\over \del \al \del \be} 
{\del t\over \del \al},
\end{eqnarray}
\begin{eqnarray}
r_{63} &=&
-{1\over 128} r_{11} {\del^3 s\over \del \al^2 \del \be} -
{1 \over 64} {\del r_{11}\over \del \al}{\del^2 s\over \del \al \del \be}-
{1 \over 128} {\del r_{11}\over \del \be}{\del^2 s\over \del \al^2} \cr
& &-{1 \over 128} {\del^2 r_{11}\over \del \al^2} {\del s \over \del \be} -
{1 \over 64} {\del^2 r_{11}\over \del \al \del \be} 
{\del s\over \del \al},
\end{eqnarray}
\begin{eqnarray}
r_{64}&=&
{C 
(-768\al^{12} - 384 \al^8 \be - 20 \al^4 - 28 \al^4 \be^2 + \be^3 - \be)
\over 2 \ Z \ \Delta^2 \ D \ r_{11}}
{\del t\over \del \al} 
\cr
& & 
+
{64 C \al^3 (8\al^4 + \be) \over \Delta^2 \ D \ r_{11}}
{\del t\over \del \be} 
+ 
{C \over \Delta \ D \ r_{11}^2}
\left( {\del t \over \del \al} {\del r_{11}\over \del \be} -
          {\del t \over \del \be} {\del r_{11}\over \del \al}
\right) 
,
\end{eqnarray}
\begin{eqnarray}
r_{65}&=&-
{C(-768\al^{12} - 384 \al^8 \be - 20 \al^4 - 28 \al^4 \be^2 + \be^3 - \be)
\over 2 \ Z  \ \Delta^2 \ D \  r_{11}}
{\del s \over \del \al} \cr
& & -{64 C \al^3 (8\al^4 + \be) \over \Delta^2 \ D \ r_{11}}
{\del s\over \del \be} 
- 
{C \over \Delta \ D \ r_{11}^2}
\left( {\del s \over \del \al} {\del r_{11}\over \del \be} -
          {\del s \over \del \be} {\del r_{11}\over \del \al}
\right) 
\end{eqnarray}
and
\begin{equation}
r_{66}=
{C \over  \Delta \  r_{11}}.
\label{app2}
\end{equation}

Next, we give some elements of $A$ and $B$.
For the elements of the first column of $A$ and $B$,
we have
\begin{equation}
A_{41}(r_{11}) = {1\over 2} \al {\del r_{11}\over \del \be} +
                 {3\over 2} \al^2 {\del^2 r_{11}\over \del \al \del \be} +
                 {1\over 2} \al^3 {\del^3 r_{11}\over \del \al^2 \del \be} -
                 {1\over 64} {\del^3 r_{11}\over \del \al^3},
\label{app3}
\end{equation}
\begin{eqnarray}
A_{61}(r_{11}) &=&
{\al \over 64 \ \Delta}r_{11} + 
{15 \al^2 \over 64 \ \Delta}
{\del r_{11}\over \del \al}+
{\al(44\al^4 + 3 \be) \over 8 \ \Delta}{\del r_{11}\over \del \be} +
{25 \al^3  \over 64 \ \Delta}{\del^2 r_{11}\over \del \al^2} \cr
& & +
{\al^2(148 \al^4 + 13 \be)
\over 8 \ \Delta}{\del^2 r_{11}\over \del \al \del \be}
+{\al^3(8\al^4 +  \be)
\over \Delta}{\del^3 r_{11}\over \del \al^2 \del \be} \cr
& & +
{1 \over 128}{\del^4 r_{11}\over \del \al^3 \del \be}
\end{eqnarray}
and
\begin{eqnarray}
B_{31}(r_{11}) &=&
-{1\over 32 \ Z}r_{11} - {3\al \over 32 \ Z}{\del r_{11}\over \del \al}
-{3 \be \over 4 \ Z}{\del r_{11}\over \del \be}
-{\al^2 \over 32 \ Z}{\del^2 r_{11}\over \del \al^2}\cr
& &-{\al\be \over 4 \ Z}{\del^2 r_{11}\over \del \al \del \be}
+{1\over 2}{\del^2 r_{11}\over \del \be^2}.
\label{app4}
\end{eqnarray}
These are proportional to the Picard-Fuchs equations (\ref{pf2para}).
$B_{51}$ and $B_{61}$ are given by
\begin{equation}
B_{51}(r_{11}) = 
{1\over Z}
\left(
{\al^2 \over 4}A_{41}(r_{11}) -
{Z \over 8} {\del \over \del \al}B_{31}(r_{11})
\right)
\label{app5}
\end{equation}
and
\begin{eqnarray}
B_{61}(r_{11}) &=&
{1\over Z}
\left\{
\left(2 \al^5 + {\al \be \over 2}\right)A_{61}(r_{11})
-{7 \al \over 32}A_{41}(r_{11})\right. \cr
& & \ \ \ \ \
\left.
-{\al^2\over 32}{\del \over \del \al}A_{41}(r_{11}) 
+{Z \over 64}{\del^2 \over \del \al^2}B_{31}(r_{11})
\right\}.
\label{app6}
\end{eqnarray}

For the elements of the second and third columns of $A$ and $B$,
we have
\begin{eqnarray}
A_{42}(r_{11},t) &=&
\left(
{\al \over 2}{\del t\over \del \be} + 
{3 \al^2 \over 2}{\del^2 t\over \del \al \del \be} +
{\al^3 \over 2}{\del^3 t\over \del \al^2 \del \be} -
{1\over 64}{\del^3 t\over \del \al^3} 
\right)r_{11} \cr
& & +
\left(
{3\al^2 \over 2}{\del t\over \del \be} +
\al^3 {\del^2 t\over \del \al \del \be} - 
{3 \over 64} {\del^2 t\over \del \al^2}
\right){\del r_{11}\over \del \al} \cr
& &+
\left(
{3\al^2 \over 2}{\del t\over \del \al} +
{\al^3 \over 2}{\del^2 t\over \del \al^2}
\right){\del r_{11}\over \del \be} +
\left(
{\al^3 \over 2}{\del t\over \del \be} -{3 \over 64}{\del t\over \del \al}
\right){\del^2 r_{11}\over \del \al^2} \cr
& &+ \al^3 {\del t\over \del \al} {\del^2 r_{11}\over \del \al \del \be},
\end{eqnarray}
\begin{eqnarray}
A_{62}(r_{11},t) &=&
{15 \ \al^2 \over 64 \ \Delta}r_{11}{\del t\over \del \al} +
{\al \ (44 \al^5 + 3 \be) \over 8 \ \Delta}r_{11}{\del t\over \del \be} +
{25 \ \al^3 \over 64 \ \Delta}r_{11}{\del ^2 t\over \del \al^2} \cr
& &
+{\al ^2 \ (148 \al^4 + 13 \be) \over 8 \ \Delta}r_{11}
{\del^2 t\over \del \al \del \be} +
{\al^3 \ (8\al^4+\be) \over \Delta}r_{11}
{\del^3 t\over \del \al^2 \del \be} \cr
& &
+{1 \over 128}r_{11}{\del^4 t\over \del \al^3 \del \be}+
{25 \ \al^3 \over 32 \ \Delta}{\del r_{11}\over \del \al}
{\del t\over \del \al} \cr
& &
+{\al^2 \ (148 \al^4 + 13 \be) \over 8 \ \Delta}
{\del r_{11}\over \del \al}{\del t\over \del \be} +
{2 \ \al^3 \ (8\al^4 + \be)\over \Delta}
{\del r_{11}\over \del \al}{\del^2 t\over \del \al \del \be}\cr
& &
+{3 \over 128}{\del r_{11}\over \del \al}
              {\del^3 t\over \del \al^2 \del \be} +
{\al^2 \ (148 \al^4 + 13 \be) \over 8 \ \Delta}
{\del r_{11}\over \del \be}{\del t\over \del \al}\cr
& &
+{\al^3 \ (8\al^4 + \be) \over \Delta}
{\del r_{11}\over \del \be}{\del^2 t\over \del \al^2}+
{1 \over 128}{\del r_{11}\over \del \be}
{\del^3 t\over \del \al^3}\cr
& & 
+{\al^3 \ (8 \al^4 + \be) \over \Delta}
{\del^2 r_{11}\over \del \al^2}
{\del t\over \del \be} +
{3 \over 128}{\del^2 r_{11}\over \del \al^2}
{\del^2 t\over \del \al \del \be} \cr
& & 
+{2 \ \al^3 \ (8 \al^4 + \be) \over \Delta}
{\del^2 r_{11}\over \del \al \del \be}
{\del t\over \del \al} +
{3 \over 128}{\del^2 r_{11}\over \del \al \del \be}
{\del^2 t\over \del \al^2}\cr
& & 
+{1\over 128}{\del^3 r_{11}\over \del \al^3}
{\del t\over \del \be} +
{3 \over 128}{\del^3 r_{11}\over \del \al^2 \del \be}
{\del t\over \del \al}
\end{eqnarray}
and
\begin{eqnarray}
B_{32}(r_{11},t) &=&
-{3 \ \al \over 32 \ Z}r_{11}{\del t\over \del \al} -
{3 \ \be \over 4 \ Z}r_{11}{\del t\over \del \be}-
{\al^2 \over 32 \ Z}r_{11}{\del^2 t\over \del \al^2}-
{\al \ \be \over 4 \ Z}r_{11}{\del^2 t\over \del \al \del \be}\cr
& &
+{1 \over 2}r_{11}{\del^2 t\over \del \be^2} -
{\al^2 \over 16 \ Z}{\del r_{11}\over \del \al}
{\del t\over \del \al} -
{\al \ \be \over 4 \ Z}{\del r_{11}\over \del \al}
{\del t\over \del \be} \cr
& &
-{\al \ \be \over 4 \ Z}{\del r_{11}\over \del \be}
{\del t\over \del \al} 
+{\del r_{11}\over \del \be}
{\del t\over \del \be}.
\end{eqnarray}
Since there exist the relations (\ref{ii1}) and (\ref{ii2}),
we only consider the elements of the second column.
$B_{52}(r_{11},t)$ and $B_{62}(r_{11},t)$ are given by
the equations which are similar to (\ref{app5}) and (\ref{app6}).

For the elements of the fourth and fifth columns of $A$ and $B$,
we have
\begin{eqnarray}
A_{44}(r_{11},t,s) &=&
{C \over D^2 \ \Delta \ r_{11}}
{\del t \over \del \al}
\left(
{\al \ (4\al^4 + \be) \over Z}
\left(
{\del t\over \del \al}{\del^2 s\over \del \al^2}
-
{\del^2 t \over \del \al^2}{\del s \over \del \al}
\right)
\right. \cr
& &
\ \ \ \ \ \ \ \ \ \ \ \ \ \ \ \ \ 
+
{64 \ \al^3}
\left(
{\del t\over \del \be} {\del^2 s\over \del \al \del \be} 
-
{\del^2 t \over \del \al \del \be}{\del s \over \del \be}
\right)
\cr
& &
\ \ \ \ \ \ \ \ \ \ \ \ \ \ \ \ \ 
+
{2}
\left(
{\del^2 t\over \del \al \del \be} {\del s\over \del \al} 
-
{\del t \over \del \al} {\del^2 s \over \del \al \del \be}
\right) \cr
& &
\ \ \ \ \ \ \ \ \ \ \ \ \ \ \ \ \ 
+{2}
\left(
{\del^2 t\over \del \al^2} {\del s\over \del \be} 
-
{\del t \over \del \be}{\del^2 s \over \del \al^2}
\right)
\cr
& &
\ \ \ \ \ \ \ \ \ \ \ \ \ \ \ \ \ 
\left.
+{2 \ D \over r_{11}}{\del r_{11}\over \del \al}
-{64 \ \al^3 \ D \over r_{11}}{\del r_{11}\over \del \be}
\right),
\end{eqnarray}
\begin{eqnarray}
B_{44}(r_{11},t,s) &=&
{C \over D^2 \ \Delta \ r_{11}}
\left(
{2}
{\del t \over \del \al} 
\left(
{\del^2 t \over \del \be^2}
{\del s \over \del \al}
-
{\del t \over \del \al}
{\del^2 s \over \del \be^2}
\right)
\right.
+
{3 \ (4 \al^4 + \be) \ D \over Z}
{\del t \over \del \al}
\cr
& &
\ \ \ \ \ \ \ \ \ \ \ \ 
+
{2}
{\del t \over \del \be}
\left(
{\del t \over \del \be}
{\del^2 s \over \del \al^2}
-
{\del^2 t \over \del \al^2}
{\del s \over \del \be}
\right)
\cr
& &
\ \ \ \ \ \ \ \ \ \ \ \ 
+
{\al^2 \ (16 \al^4 \be + 3 \be^2 +1) \over 8 \ Z^2}
{\del t \over \del \al}
\left(
{\del t \over \del \al}
{\del^2 s \over \del \al^2} 
-
{\del^2 t \over \del \al^2}
{\del s \over \del \al}
\right)
\cr
& &
\ \ \ \ \ \ \ \ \ \ \ \ 
+
{\al \ (4 \al^4 + \be) \over Z}
{\del t \over \del \be}
\left(
{\del^2 t \over \del \al^2}
{\del s \over \del \al}
-
{\del t \over \del \al}
{\del^2 s \over \del \al^2}
\right)
\cr
& &
\ \ \ \ \ \ \ \ \ \ \ \ 
+
{64 \ \al^3}
{\del t \over \del \be}
\left(
{\del^2 t \over \del \al \del \be}
{\del s \over \del \be}
-
{\del t \over \del \be}
{\del^2 s \over \del \al \del \be}
\right)
\cr
& &
\ \ \ \ \ \ \ \ \ \ \ \ 
+
{2 \ \al \ (4\al^4 + \be) \ D \over Z \ r_{11}}
{\del t \over \del \al}
{\del r_{11} \over \del \al} 
-
{2 \ D \over r_{11}}
{\del t \over \del \be}
{\del r_{11} \over \del \al}
\cr
& &
\ \ \ \ \ \ \ \ \ \ \ \ 
-
{4 \ D \over r_{11}}
{\del t \over \del \al}
{\del r_{11} \over \del \be} 
+
{64 \al^3}
{\del t \over \del \al}
\left(
{\del t \over \del \be}
{\del^2 s \over \del \be^2} 
-
{\del^2 t \over \del \be^2}
{\del s \over \del \be}
\right)
\cr
& &
\ \ \ \ \ \ \ \ \ \ \ \ 
+
{2}
{\del t \over \del \be}
\left(
{\del t \over \del \al}
{\del^2 s \over \del \al \del \be}
-
{\del^2 t \over \del \al \del \be}
{\del s \over \del \al}
\right)
+
{64 \ \al^3 \ D \over r_{11}}
{\del t \over \del \be}
{\del r_{11} \over \del \be}
\cr
& &
\ \ \ \ \ \ \ \ \ \ \ \ 
\left.
+
{\al \ (4 \al^4 + \be) \over Z}
{\del t \over \del \al}
\left(
{\del^2 t \over \del \al^2}
{\del s \over \del \be}
-
{\del t \over \del \be}
{\del^2 s \over \del \al^2}
\right)
\right)
\end{eqnarray}
and
\begin{eqnarray}
B_{54}(r_{11},t,s) &=&
{C \over D^2 \ \Delta \ r_{11}}
{\del t \over \del \al}
\left(
{\al^2 \ (16 \al^4 \be + 3 \be^2 + 1) \over 2 \ Z^2}
\left(
{\del t \over \del \al}
{\del^2 s \over \del \al \del \be} 
-
{\del^2 t \over \del \al \del \be}
{\del s \over \del \al}
\right)
\right.
\cr
& &
\ \ \ \ \ \ \ \ \ \ \ \ \ \ \ \ 
+
{8}
\left(
{\del t \over \del \be} 
{\del^2 s \over \del \be^2} 
-
{\del^2 t \over \del \be^2}
{\del s \over \del \be}
\right)
\cr
& &
\ \ \ \ \ \ \ \ \ \ \ \ \ \ \ \ 
+
{4 \ \al \ (4 \al^4 + \be) \over Z}
\left(
{\del^2 t \over \del \al \del \be}
{\del s \over \del \be} 
-
{\del t \over \del \be}
{\del^2 s \over \del \al \del \be}
\right)
\cr
& &
\ \ \ \ \ \ \ \ \ \ \ \ \ \ \ \ 
+
{4 \ \al \ (4 \al^4 + \be) \over Z}
\left(
{\del^2 t \over \del \be^2}
{\del s \over \del \al} 
-
{\del t \over \del \al}
{\del^2 s \over \del \be^2}
\right)
\cr
& &
\ \ \ \ \ \ \ \ \ \ \ \ \ \ \ \ 
+
{3 \ \al \ (16 \al^4 \be + 3 \be^2 + 1) \over 2 \ Z^2}D
\cr
& &
\ \ \ \ \ \ \ \ \ \ \ \ \ \ \ \ 
+
{\al^2 \ (16 \al^4 \be + 3 \be^2 + 1) \ D \over 2 \ Z^2 \ r_{11}}
{\del r_{11} \over \del \al}
 \cr
& &
\ \ \ \ \ \ \ \ \ \ \ \ \ \ \ \ 
\left.
-
{4 \ \al \ (4 \al^4 + \be) \ D \over Z \ r_{11}}
{\del r_{11} \over \del \be}
\right).
\end{eqnarray}
The other elements contain too many terms to be reproduced here 
explicitly.
For example, $A_{64}(r_{11},t,s)$ and  $B_{64}(r_{11},t,s)$ possess
328 and 532 terms, respectively,
and hence we shall refrain from presenting them.

Finally, let us discuss some properties of  
the solutions of (\ref{*2}).
If we write 
\begin{equation}
A_{44}(r_{11},t,s) = {C \over D^2 \ \Delta \ r_{11}}
                     {\del t \over \del \al}F(r_{11},t,s)
\end{equation}
and
\begin{equation}
B_{54}(r_{11},t,s) = {C \over D^2 \ \Delta \ r_{11}}
                     {\del t \over \del \al}E(r_{11},t,s),
\end{equation}
then $F(r_{11},t,s)$ and $E(r_{11},t,s)$ are bilinear 
and symmetric in $t$ and $s$
\begin{equation}
F(r_{11},t,s) = - F(r_{11},s,t),
\ \ \  
E(r_{11},t,s) = - E(r_{11},s,t).
\end{equation}
It is thus easy to show that 
the solutions of $A_{44}=0$ and $B_{54} = 0$ are allowed to be 
transformed
linearly; $t \rightarrow {\tilde t} = a t + b s$ and 
$s \rightarrow {\tilde s} = c t + d s$. 
However, in spite of the non-linear property
of $B_{44}$, $A_{64}$ and $B_{64}$,
we have observed by explicit calculations that
the solutions of $B_{44}=0$, $A_{64}=0$ and $B_{64}=0$
are also allowed to be transformed linearly. This implies that 
these equations may reduce 
to more fundamental ones.

\newpage

%
%

\end{document}